\documentclass[a4paper,11pt]{article}
\usepackage{amsmath}
\usepackage{jheppub}
\usepackage{amssymb}
\usepackage{graphicx}
\usepackage[T1]{fontenc}
\usepackage{slashed}
\usepackage[utf8]{inputenc}
\usepackage{xspace}
\usepackage{color}
\usepackage{ulem}
\usepackage{array}
\usepackage{ulem,fancyvrb}
\usepackage{xcolor}

\begin{document}

\title{The Light Sgoldstino Phenomenology: Explanations for the Muon $(g-2)$ Deviation and KOTO Anomaly}
\date{\today}
\author[1]{Xuewen Liu,}
\author[1,3]{Ying Li, }
\author[4,5]{Tianjun Li,}
\author[1,2]{Bin Zhu}

\affiliation[1]{Department of Physics, Yantai University, Yantai 264005, China}
\affiliation[2]{Department of Physics, Chung-Ang University, Seoul 06974, Korea}
\affiliation[3]{Center for High Energy Physics, Peking University, Beijing 100871, China}
\affiliation[4]{CAS Key Laboratory of Theoretical Physics, Institute of Theoretical Physics,
Chinese Academy of Sciences, Beijing 100190, China}
\affiliation[5]{School of Physical Sciences, University of Chinese Academy of Sciences,
Beijing 100049, China}
\emailAdd{xuewenliu@ytu.edu.cn; liying@ytu.edu.cn; tli@itp.ac.cn; zhubin@mail.nankai.edu.cn}

\abstract{
In this work, we study the long-standing experimental anomaly in muon $(g-2)$ and also recent anomalous excess in $K_L\to \pi^0+\nu\bar\nu$ at the J-PARC KOTO experiment with sgoldstino. After supersymmetry breaking, the interactions between quarks and sgoldstino ($s$) make the decays $K\to \pi+s$ sizable through loop diagrams, which affects the measurements of decays $K\to \pi+\mathrm{invisible}$. Furthermore, the couplings between photons and sgoldstino contribute to $\Delta a_\mu$ as well as the bino-slepton contribution.  With satisfying all known experimental constraints such as from NA62, E949, E137, Orsay, KTEV and CHARM experiments, these two anomalies can be explained simultaneously. The mass of CP-even sgoldstino is close to the neutral pion mass which does not violate the Grossman-Nir bound. The parameter space can be further tested in future NA62, DUNE experiments, as well as experiments in the LHC.
}
\maketitle

\section{Introduction}
After the Higgs particle was discovered at the Large Hadron Collider (LHC)~\cite{Aad:2012tfa, Chatrchyan:2012ufa}, the Standard Model (SM) has been confirmed as a  successful low energy description of Nature. Although most of the SM predictions are consistent with the experimental data, there still exists some problems in the SM. For example, gauge hierarchy problem, gauge coupling unification, dark matter, baryon asymmetry, and neutrino masses and mixings, etc. Supersymmetry provides a natural solution to the gauge hierarchy problem. In the Supersymmetric SMs (SSMs) with R-parity, gauge coupling unification can be achieved, the lightest supersymmetric particle (LSP) can be a dark matter candidate, and the electroweak gauge symmetry can be broken radiatively due to large top quark Yukawa coupling, etc. Therefore, supersymmetry is a promising scenario for new physics beyond the SM. However, we have strong constraints on the parameter space in the SSMs from the supersymmetry searches at the LHC. The interesting question is whether there exists some light particles in the SSMs which can be probed or can explain the anomalies at the current experiments. Therefore we can probe supersymmetry indirectly.

We shall present one example in this paper. Once supersymmetry is broken spontaneously, we have a Goldstone fermion ${\widetilde G}$, i.e. goldstino. The superpartner of goldstino is called  sgoldstino $S=\frac{1}{\sqrt{2}}(s+ia)$ where $s$ and $a$ are a CP-even and CP-odd real scalars. In particular, in the low energy supersymmetry breaking, i.e. $N=1$ supersymmetry is broken at low energy around TeV scale~\cite{Dudas:2012fa,Astapov:2014mea,Petersson:2015mkr,Ding:2016udc,Baratella:2016daa,Demidov:2016dye,Co:2017pyf,Demidov:2020jne}, sgoldstino is so light that it can be  probed in low energy experiments. Therefore, we would like to explore the sgoldstino phenomenology.

Experimentally, there exists a $3.7~\sigma$ discrepancy for the anomalous magnetic moment of the muon $a_\mu = (g_\mu-2)/2$ between the experimental results~\cite{Bennett:2006fi, Tanabashi:2018oca} and theoretical predictions~\cite{Davier:2017zfy, Blum:2018mom, Keshavarzi:2018mgv, Davier:2019can, Aoyama:2020ynm}
\begin{equation}
\label{muon} \Delta a_\mu = a_\mu^{\rm exp}-a_\mu^{\rm th} = (2.74\pm 0.73)\times 10^{-9}~,~\,
\end{equation}
which is a well-known long-standing deviation. Computing the hadronic light-by-light contribution with all errors under control by using lattice QCD, several groups are trying to improve the precision of the SM predictions~\cite{Aubin:2019usy, Blum:2015you, Lehner:2019wvv, Davies:2019efs, Borsanyi:2020mff}. The recent lattice calculation for  the hadronic light-by-light scattering contribution has confirmed the $\Delta a_\mu$ discrepancy~\cite{Blum:2019ugy}, and then a new physics explanation of the discrepancy is needed. Also, the ongoing experiment at Fermilab~\cite{Grange:2015fou, Fienberg:2019ddu} and one planned at J-PARC~\cite{Saito:2012zz} will try to reduce the uncertainty.

In addition, the flavor changing processes like rare $K$ meson decays, $K_L \rightarrow \pi^0 \nu \bar{\nu}$ and $K^+ \rightarrow \pi^+ \nu \bar{\nu}$, which are loop suppressed in the SM~\cite{Littenberg:1989ix, Cirigliano:2011ny}, are very sensitive to the new physics beyond the SM~\cite{Buras:2015qea, Tanimoto:2016yfy, Crivellin:2017gks, Bordone:2017lsy, Endo:2017ums, He:2018uey, Chen:2018ytc}. The SM predictions are~\cite{Buras:2015qea}
\begin{eqnarray} \label{neutralKSM}
\text{Br}(K_L \rightarrow \pi^0 \nu \bar{\nu})_{\text{SM}} &=& (3.00 \pm 0.30) \times 10^{-11} ~,~\, \\ \label{chargedKSM}
\text{Br}(K^+ \rightarrow \pi^+ \nu \bar{\nu})_{\text{SM}} &=& (9.11 \pm 0.72) \times 10^{-11}~.~\,
\end{eqnarray}
These processes are studied at the KOTO experiment~\cite{Ahn:2018mvc, Yamanaka:2012yma} at J-PARC~\cite{Nagamiya:2012tma} and NA62 experiment~\cite{NA62:2017rwk} at CERN. In particular, four candidate events have been observed in the signal region of {$K_L \rightarrow \pi^0 \nu \bar{\nu}$} search at the KOTO experiment, while the SM prediction is only $0.10 \pm 0.02$. One event can be suspected as a background coming from the SM upstream activity, and the other three can be considered as signals since they are not consistent with the currently known background. Note that the single event sensitivity is $6.9 \times 10^{-10}$, three events are consistent with

\begin{equation} \label{koto19}
    {\mathrm{Br}(K_{L} \rightarrow  \pi^0 \nu \bar{\nu} )_{{_{\rm KOTO19}}}
= 2.1^{+2.0(+4.1)}_{-1.1(-1.7)} \times 10^{-9},}
\end{equation}
at 68(90)$\%$ confidence {level} (C.L.), including statistical uncertainties, whose central value is almost two orders of magnitude larger than the SM prediction. This new result includes the interpretation of photons and invisible final states as $\nu \bar{\nu}$ and is in agreement with their previous bounds~\cite{Ahn:2018mvc} 
\begin{equation}
\text{Br}(K_{L} \rightarrow  \pi^0 \nu \bar{\nu} )_{\rm KOTO18}< 3.0 \times 10^{-9}~.~\,
\end{equation}
However the charged kaon decay searches have not observed any excess events. The recent update from NA62 puts a bound
 \begin{equation}
\text{Br}(K^+ \rightarrow  \pi^+ \nu \bar{\nu} )_{\rm NA62}< 2.44 \times 10^{-10}
\end{equation}
at 95$\%$ C.L.,  which is consistent with the SM prediction of Eq.~(\ref{chargedKSM}).

Furthermore, the generic neutral and charged kaon decays satisfy the following Grossman-Nir (GN) bound~\cite{Grossman:1997sk}
\begin{equation}
{\rm Br}\left(K_L \rightarrow \pi^0 \nu {\bar \nu}\right)  \leq 4.3 \times \left(K^+ \rightarrow \pi^+ \nu {\bar \nu}\right) ~,~\, \label{GNB}
\end{equation}
which depends on the isospin symmetry and kaon lifetimes. Because the explanations for the KOTO anomaly might be strongly constrained by the GN bound, the new physics explanation for the KOTO anomaly is required to not only generate three anomalous events, but also satisfy the GN bound. Recently, the  KOTO anomaly has been studied extensively  in the literatures~\cite{Kitahara:2019lws,Ballett:2019pyw, Mandal:2019gff,  Fabbrichesi:2019bmo, Egana-Ugrinovic:2019wzj, Dev:2019hho, Li:2019fhz, Jho:2020jsa, Liu:2020qgx, MartinCamalich:2020dfe, Yamaguchi:2020eub,Ertas:2020xcc, Banerjee:2020kww,Dev:2020eam, Liao:2020boe, Cline:2020mdt, Ziegler:2020ize, Gori:2020xvq,He:2020jzn, He:2020jly, Datta:2020auq, Dutta:2020scq,Lichard:2020lul,Haghighat:2020nuh,Altmannshofer:2020pjb}.

In this paper, we shall explain the muon anomalous magnetic moment and KOTO anomaly simultaneously via a light sgoldstino. This paper is structured as follows: In Sec. \ref{sec:motivation}, we review the basic motivation of sgoldstino and introduce relevant interaction between sgoldstino and SM particles. In terms of them, we can analysis the sgoldstino phenomenology quantitatively. Sec. \ref{sec:signal} {investigates} how we can fit the muon $(g-2)$ and KOTO anomaly simultaneously. In addition, all the relevant constraints are considered seriously. Finally, we make a summary in Sec. \ref{sec:conclusion}.

\section{Motivation of Sgoldstino and its implication on phenomenology}
\label{sec:motivation}

Once supersymmetry is spontaneously broken, there must exists a massless goldstino $\widetilde G$~\cite{Martin:1997ns}.  Goldstino is a Goldstone  fermion and becomes the longitudinal component of gravitino with its mass being lifted by gravity correction i.e. $m_{3/2}\sim F/M_P$. Thus, the existence of gravitino is an inevitable prediction of local supersymmetry from super-Higgs mechanism. The goldstino chiral superfield is written as
\begin{equation}
\Phi~=~ S+\sqrt{2} \theta {\widetilde G}+F \theta^{2}~.
\end{equation}
Generally we assume that supersymmetry is broken by the dynamics of Goldstino superfield $\Phi$ and $F$ encodes the SUSY breaking information. Through some weak couplings of Goldstino to those of the MSSM fields, some non-renormalizable operator gives rise to soft breaking terms.  For example, any supersymmetric theory contains operator i.e. $\mathcal{L}_{\mathrm{eff}}$=$M_{a}/F \int d^{2} \theta \Phi W_{a}^{\alpha} W_{\alpha}^{a}$, which is used to generate non-vanishing gaugino mass in various SUSY breaking and mediation mechanism, 
\begin{equation}
\mathcal{L}_{\mathrm{eff}}=\frac{M_{a}}{2 F} \int d^{2} \theta \Phi W_{a}^{\alpha} W_{\alpha}^{a}=\frac{M_{a}}{2} \lambda_{a} \lambda_{a}+\frac{M_{a}}{2 \sqrt{2} F}\left(s F_{a}^{\mu \nu} F_{\mu \nu}^{a}-a F_{a}^{\mu \nu} \tilde{F}_{\mu \nu}^{a}\right). 
\label{eqn:gaugino}
\end{equation}

The reason the operator is non-renormalizable comes from super-trace theorem. We should mention that, the operator not only generates gaugino soft masses but includes an inevitable coupling between sgoldstino and SM gauge bosons. The generic Lagrangian of our model is thus given by
\begin{eqnarray}
  -{\mathcal{L}}&& \supset \frac{M_{i}}{2} \lambda_{i} \lambda_{i}
+ \frac{M_{i}}{2\sqrt{2}F} F_{\mu\nu}^{i}
  \left(-sF^{\mu\nu i}+ia {\tilde F}^{\mu\nu i}\right) \nonumber \\
&&+\left[\frac{S}{\sqrt{2}F}(A_{ij}^{U}Q_{i}U_{j}^{c}H_{u}+ A_{ij}^{D}Q_{i}D_{j}^{c}H_{d}+A_{ij}^E L_i E_j^c H_d)
+ {\rm H.C.} \right]~,~
\label{lagrangian}
\end{eqnarray}
where $M_i$  are gaugino masses from $F$-term of $\Phi$, and $\tilde F_{\mu\nu} = \frac{1}{2}\epsilon_{\mu\nu\alpha\beta}F^{\alpha\beta}$. Also, $A_{ij}^{U,D,E}$ are the couplings, but not necessary the same as the Yukawa couplings in the SM in general. Also, $Q_{i},~U_{i}^{c},$, $D_{i}^{c}$, $L_i$, $E_i^c$, $H_u$, and $H_d$ are the left-handed quark doublets, right-handed up-type quarks, and right-handed down-type quarks, left-handed lepton doublets, right-handed charged leptons,up-type and down-type Higgs doublets, respectively.

The mass of sgoldstino arises from higher order K\"ahler potential. To obtain the mass splitting between $s$ and $a$, we consider the  high-dimensional K\"ahler potential as follows~\cite{Ding:2016udc}
\begin{align}
\kappa =\kappa_{1} \frac{(S\bar S)^{2}}{M^2}+\left[\kappa_{2} \frac{S\bar S^{3}}{2M^2}+ \kappa_{2}^{*} \frac{S^{3}\bar S}{2M^2}\right]
\end{align}
And the scalar potential becomes~\cite{Ding:2016udc}
\begin{align}
V=\kappa_{1} \frac{|F_{S}|^{2}}{M^2}|S|^{2}+\left[\kappa_{2} \frac{|F_{S}|^{2}}{2M^2} \bar S^{2}+ \kappa_{2}^{*} \frac{|F_{S}|^{2}}{2M^2}{S}^{2}\right].
\label{scalar_pot}
\end{align}
Taking $\kappa_{2}$ real ($\kappa_{2}=\kappa_{2}^{*}$), $m_{S}^{2}=\frac{|F_{S}|^{2}}{M^{2}}$, and using $S=\frac{1}{\sqrt{2}}(s+ia)$, we can rewrite Eq.~(\ref{scalar_pot}) as below~\cite{Ding:2016udc}
\begin{align}
V=(\kappa_{1}+\kappa_{2})\frac{m_{S}^2}{2}s^{2} + (\kappa_{1}-\kappa_{2})\frac{m_{S}^2}{2}a^{2}.
\label{scalar_pot2}
\end{align}
Thus, there are two simple cases for mass splitting: 

{\bf Case (1):} Single resonance.
When $\kappa_{1}\simeq \kappa_{2}$, we have $m_{s}\gg m_{a}$.  

{\bf Case (2):} Twin resonances.
When $\kappa_{1}\gg \kappa_{2}$, then $m_{s}\simeq m_{a}$.

In this paper, we shall consider {\bf Case (1)}, and treat $a$ nearly massless in our setup. Therefore, the light scalar to mimic KOTO anomaly is the real component of sgoldsitino, $s$.  Besides, there is a tension between the neutral and charged sector which can be used to fix our sgoldstino mass. If we believe that these decays are dominated by the transitions with isospin $\Delta I=\frac{1}{2}$,  these two decays are related by the GN bound \cite{Grossman:1997sk}, as shown in Eq.(\ref{GNB}). The numerical factor 4.7 comes from the differences in widths, isospin breaking effects and QED corrections. This bound puts very strong constraints on any explanations of the KOTO anomaly with new physics, since both two decays are induced by the same transition {$s~{\rm quark} \to d~{\rm quark}+s$}. Recent studies based on the effective theory showed that a violation of GN bound by the new physics contribution is quite nontrivial \cite{He:2020jzn,He:2020jly}. Considering the experimental sensitivities of the charged and neutral Kaon experiments, the KOTO anomaly can be explained without violating GN bound if a new scalar with a mass of about pion is stable or  with a lifetime lower than about a nanosecond, which has been pointed out in \cite{Fuyuto:2014cya}. Very recently,  the authors  in \cite{Kitahara:2019lws} pointed out that a light scalar, with mass different from the pion mass and with a long lifetime, but not necessarily stable, can also explain the observed KOTO excess. In short, the mass of the sgoldstino should be close to the mass of pion i.e. $m_s\in [50, 200]\mathrm{MeV}$.

The additional interaction for sgoldstino and SM particles in Eq.~(\ref{lagrangian}) can be ignored at high scale SUSY breaking, since it is suppressed by SUSY breaking scale  $\sqrt{F}$. But it actually provides large deviation from SM when SUSY breaking scale is low enough such as $\sqrt{F}\sim [10^3, 10^5]\mathrm{GeV}$. So the question becomes whether or not we can fit the signal of KOTO anomaly, muon $(g-2)$ and evade the current bounds for light sgoldstino and sparticles. The relevant interactions between sgoldstino and the SM particles that are responsible for these anomalies can be obtained from Eq.(\ref{lagrangian}),
\begin{eqnarray}
-\mathcal{L}_{\mathrm{eff}}&=&\frac{M_{\gamma}}{2 \sqrt{2} F} s F^{\mu \nu} F_{\mu \nu}+\frac{M_{3}}{2 \sqrt{2} F} s G^{a \mu \nu} G_{\mu \nu}^{a} \nonumber \\
&& + \left(\frac{A_{ij}^U v}{\sqrt{2}F} s Q_{i}U_{i}^{c} + \frac{A_{ij}^D v}{\sqrt{2}F} s Q_{i}D_{j}^{c}
+  \frac{A_{ij}^E v}{\sqrt{2}F} s L_{i}E_{j}^{c} +{\rm H.C.} \right)~,~\,
\label{eqn:eff}
\end{eqnarray}
where $M_{\gamma}$ is the combination of $M_1$ and $M_2$ after electroweak symmetry breaking, i.e.
 $M_{\gamma}=M_1\cos^2\theta_w+M_2\sin^2\theta_w$. It determines the interaction strength between sgoldstino and photon.  There are several points that we should mention on the relevant lagrangian in Eq.(\ref{eqn:eff}):
\begin{itemize}
\item For naturalness, we can assume the universal trilinear soft terms as $A_{ij}^U= A_{ij}^D = A_{ij}^E  \equiv A_0\delta_{ij} $. Therefore, the trilinear soft terms do not lead to any tree-level flavor-changing processes. And then sgoldstino shares the same interaction strength for both quarks and leptons, which plays a crucial role in our phenomenological study,
\begin{equation}
\lambda_q=\lambda_l=\frac{A_0 v}{\sqrt{2}F}~,~\,
\end{equation}
where we denote $\lambda_q$ and $\lambda_l$ as the effective coupling between sgoldstino and quarks as well as leptons, respectively. Generically, $\lambda_q$ is around $10^{-2}$ to obtain the correct neutral Kaon decay. Meanwhile, such a tiny $\lambda_l$ can not generate $\Delta a_{\mu}$ at the required order $10^{-9}$. Here, we will not consider the non-universal trilinear soft terms as a solution since it requires about 1\% fine-tuning. So this becomes a challenge for sgoldsino to explain the two anomalies at the same time. We will figure out how to solve the problem in next section.

\item
Sgoldstino can also couple to W-boson and Z-boson through gaugino mass $M_1$ and $M_2$. But these two couplings cannot affect the Kaon decay. Of course, the interaction with W-boson provides additional channel for sgoldstino decay into photon. However, it is too small compared with the tree-level decay induced by $M_{\gamma}$.

\item MeV sgoldstino can not decay into gluon pair and quarks since it does not have enough energy to hadronization. Thus, even gluino is much heavier than bino, wino, it cannot decay into gluon pair channel.

\item Coupling between $s$ and $a$ can be induced by higher order Kahler potential. But these couplings are highly suppressed by UV cut-off $M$. It is acceptable that $s$ has no coupling with $a$ effectively.
\end{itemize}

\begin{figure}[!htbp]
	\centering
	\includegraphics[width=0.5\textwidth]{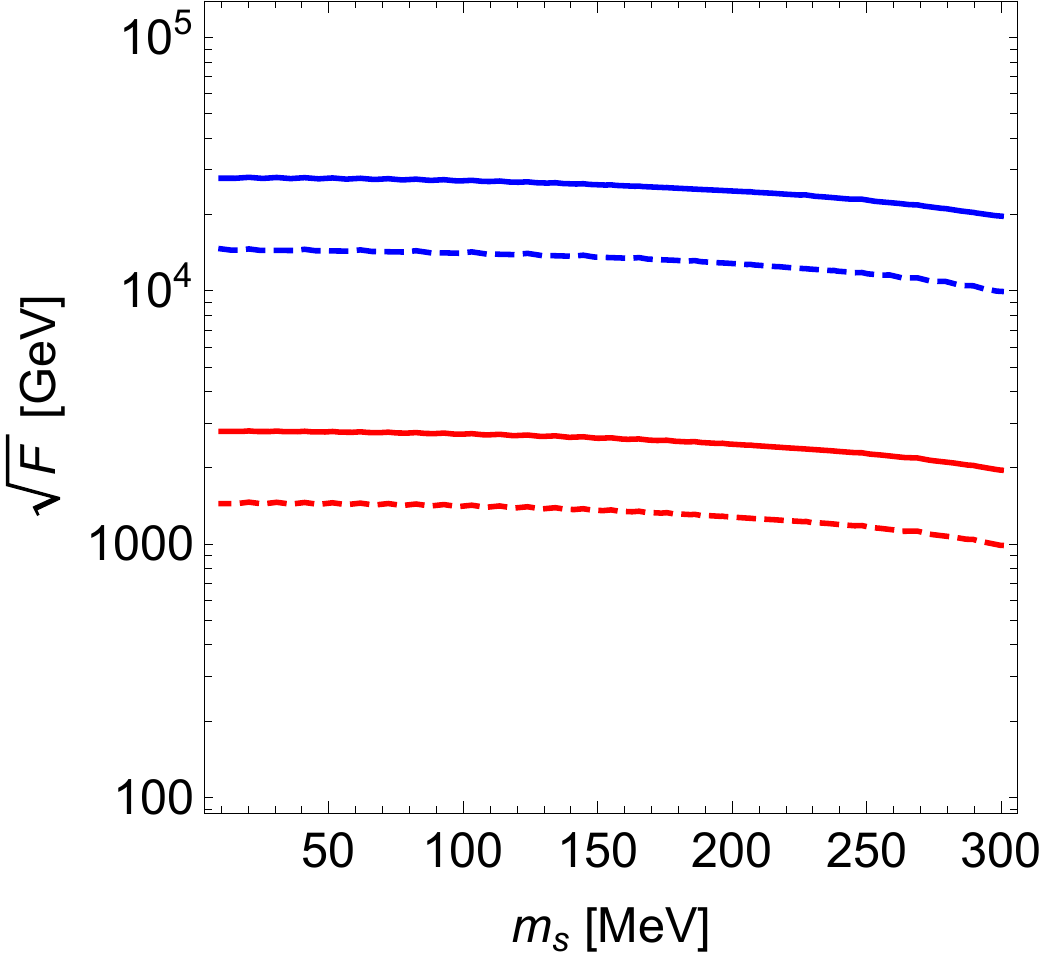}
	\caption{Contours for branching ratios of processes
	$K_L\to\pi^0 s$ and $K^+\to\pi^+ s$.
	The solid colored lines indicate the contours for BR($K_L\to\pi^0 s$).
	Dotted lines of corresponding colors show where BR($K^+\to\pi^+ s$)
	achieves the corresponding value: blue for $\mathrm{BR}=10^{-8}$, red for $\mathrm{BR}=10^{-6}$.
	{We set $A_0=0.2 \sqrt F$ here.}
	}
	\label{fig:Kaon}
	\end{figure}

With Eq.(\ref{eqn:eff}) at hand, we can easily calculate the neutral Kaon $K_L$ decay widths
\begin{align}
\Gamma\left(K_{L} \rightarrow \pi^{0} s\right) &=\frac{\left(\operatorname{Re}\left[g\left(\lambda_{q}\right)\right]\right)^{2}}{16 \pi m_{K}^{3}} \lambda^{1 / 2}\left(m_{K}^{2}, m_{\pi}^{2}, m_{s}^{2}\right),  \\
g\left(\lambda_{q}\right) &=\frac{3 m_{K}^{2}}{32 \pi^{2} v^{3}} \lambda_{q} f_{+}(0) \sum_{q=u, c, t} m_{q}^{2} V_{q d}^{*} V_{q s}, 
\end{align}
with the K\"all\'en function
\begin{equation}
\lambda(x, y, z)=x^{2}+y^{2}+z^{2}-2 x y-2 y z-2 x z .
\end{equation}
The form factor $f_{+}(0)$ is set to be 0.9709 by lattice QCD. The charged Kaon decay can be obtained by replacing $\operatorname{Re}[g(\lambda_q)]$ to $\left|g(\lambda_q)\right|$ and corresponding mass parameters. In figure \ref{fig:Kaon},  we show the contours for BR($K_L\to\pi^0 s$) and BR($K^+\to\pi^+ s$) (solid and {dashed} lines respectively) in the $(m_s- \sqrt F)$ plane.  We find that in the range of sgoldstino {mass},  the branching ratio is fairly insensitive to $m_s$ and thus determined mostly by $\sqrt F$, where a larger $\sqrt F$ corresponds to a smaller branching ratio.  We can also find that for the same branching ratio for both processes, BR($K^+\to\pi^+ s$) requires a bigger value of  $\sqrt F$, which also means that for a given set of parameters, the values of BR($K^+\to\pi^+ s$) is smaller than that of BR($K_L\to\pi^0 s$). It naturally explains why we only observe the neutral Kaon decay in KOTO experiment.

\begin{figure}[!htbp]
	\centering
	\includegraphics[width=0.48\textwidth]{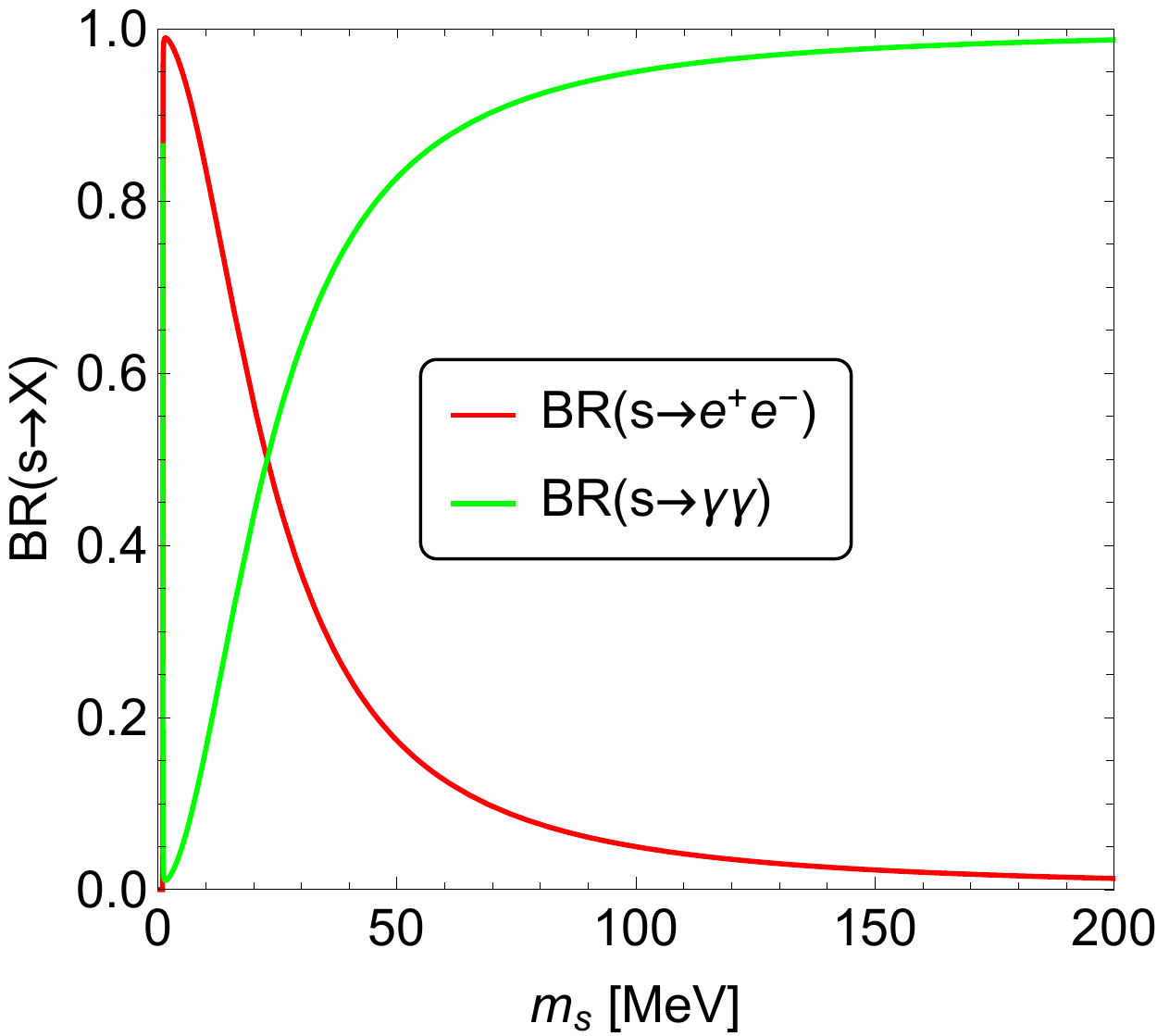}
	\caption{Branching ratios of the allowed $s$ decay modes as function of scalar mass $m_s$,
	with typical values of SUSY parameters ($M_1=100$ GeV, $M_2=1$ TeV, $\sqrt F=5\times10^4$ GeV, $A_0/\sqrt F=0.2$).
	The red curve shows branching ratios BR$(s\to e^+e^-)$,
	and the green curve represents the BR$(s\to \gamma\gamma)$.}
	\label{fig:s-decay}
	\end{figure}

For the decay of $s$, we have two different channels in the mass range that we interested in: decay into photon pair and decay into electron pair, and the decay widths are
\begin{align}
&\Gamma(s \rightarrow \gamma \gamma)=\frac{1}{32 \pi}\left(\frac{M_{\gamma}^{2}}{F^{2}}\right) m_{s}^{3},\\
&\Gamma(s \rightarrow e^+ e^-)=\frac{A_{0}^{2} m_{e}^{2} m_{s}}{16 \pi^{2} F^{2}}\left(1-\tau_{e}\right)^{\frac{3}{2}},
\end{align}
where $\tau_e=4m_e^2/m_s^2$. In figure \ref{fig:s-decay}, we present the branching ratios of the possible decay modes of $s$, i.e. $s\to e^+e^-$ and $s\to \gamma\gamma$. Since the diphoton width receives tree level contributions from $M_1, M_2$, one can see that this process will  be the dominant decay mode with the scalar mass larger than tens MeV in figure~\ref{fig:s-decay}.

\section{Signal of $(g-2)_{\mu}$, KOTO anomaly and Constraints}\label{sec:signal}

\begin{figure}[!htbp]
	\centering
	\includegraphics[width=0.6\textwidth]{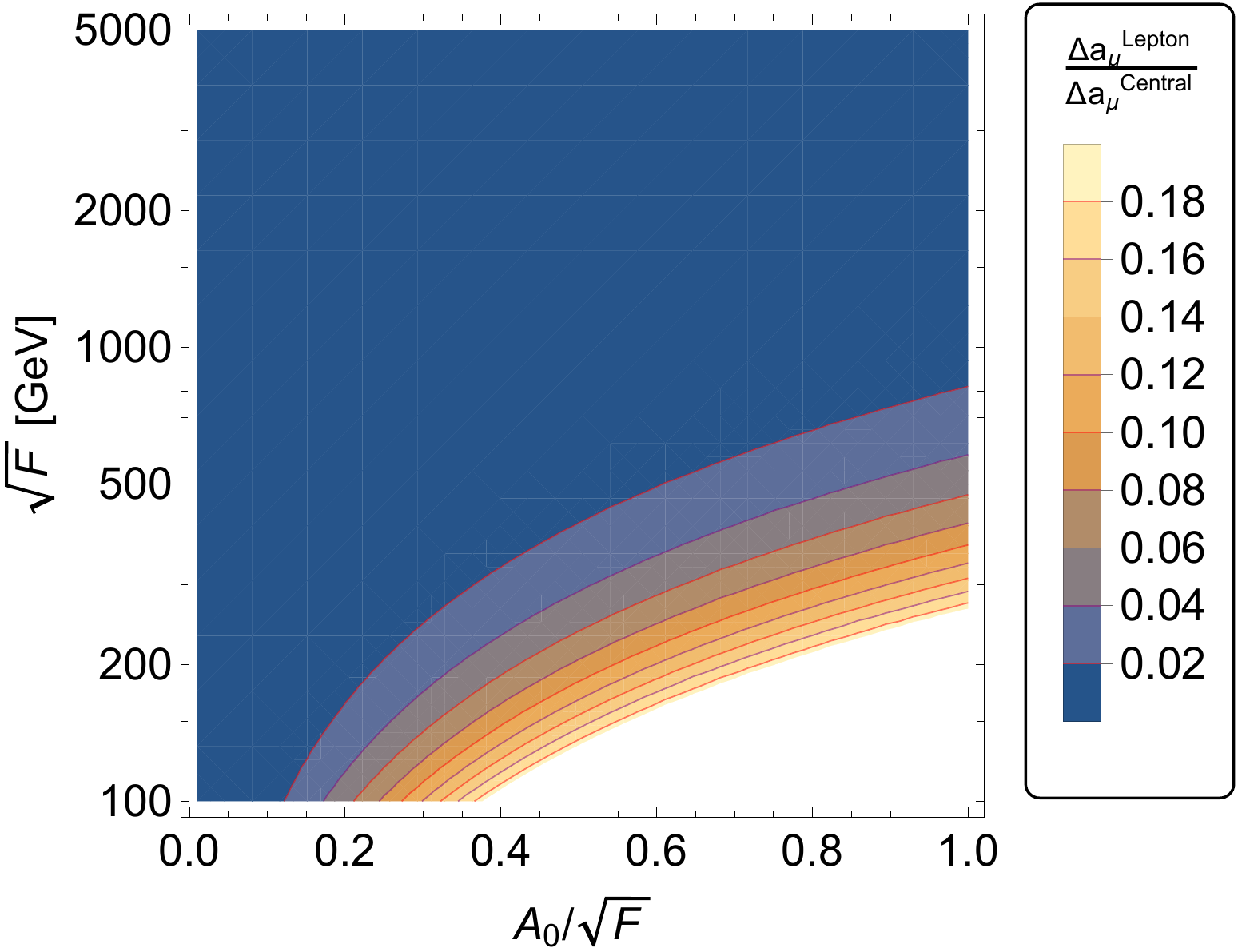}
	\caption{Contours of sgoldstino contribution to muon $g-2$ in the parameter space ($A_0/\sqrt F, \sqrt F$). Here we choose $m_s=100$ MeV for illustration.
	{$\Delta a_{\mu}^{\mathrm{Lepton}}$ and $\Delta a_{\mu}^{\rm central}$ stand for the contribution of sgoldstino and the central value of  $\Delta a_{\mu}$ repectively.}}
	\label{fig:muon1}
	\end{figure}

In this section, we focus on how to use sgoldstino to explain KOTO anomaly and muon $(g-2)$ simultaneously. 
Naively, we can estimate the contributions to muon $(g-2)$ from sgoldstino, since it couples to SM leptons with effective coupling $\lambda_l$. Due to the fact that the sgoldstino-lepton coupling is equivalent to that with quarks, its contribution is very small when we impose KOTO anomaly requirement, see figure \ref{fig:muon1}. Here we define the contribution from $\lambda_l=A_0 v/\sqrt{2}F$ to be $\Delta a_{\mu}^{\mathrm{Lepton}}$. The ratio between $\Delta a_{\mu}^{\mathrm{Lepton}}$ and central value of required $\Delta a_{\mu}$ is smaller than $0.2$ in all the parameter space. That is to say, only $\lambda_l$ contribution is not enough to generate required muon $(g-2)$.
However there always exist contributions coming from neutralino and sleptons in the SUSY framework. Here we include all the five important one-loop diragrams: charigino-sneutrino loop ($\mathrm{C}$), wino-slepton loop ($\mathrm{W}$), bino-slepton loop ($\mathrm{B}$), bino-higgsino loop ($\mathrm{BHR}$ and $\mathrm{BHL}$) \cite{Martin:2001st,Abdughani:2019wai},

\begin{equation}
\begin{aligned}
&\Delta a_{\mu}^{\mathrm{C}}=\frac{g_{2}^{2} m_{\mu}^{2}}{8 \pi^{2}} \frac{M_{2} \mu \tan \beta}{m_{\widetilde{\nu}_{\mu}}^{4}} F_{a}\left(\frac{M_{2}}{m_{\widetilde{\nu}_{\mu}}}, \frac{\mu}{m_{\widetilde{\nu}_{\mu}}}\right), \\
&\Delta a_{\mu}^{\mathrm{W}}=-\frac{g_{2}^{2} m_{\mu}^{2}}{16 \pi^{2}} \frac{M_{2} \mu \tan \beta}{m_{\widetilde{\mu}_{\mathrm{L}}}^{4}} F_{b}\left(\frac{M_{2}}{m_{\widetilde{\mu}_{\mathrm{L}}}}, \frac{\mu}{m_{\widetilde{\mu}_{\mathrm{L}}}}\right), \\
&\Delta a_{\mu}^{\mathrm{B}}=\frac{g_{Y}^{2} m_{\mu}^{2}}{8 \pi^{2}} \frac{\mu \tan \beta}{M_{1}^{3}} F_{b}\left(\frac{m_{\widetilde{\mu}_{\mathrm{L}}}}{M_{1}}, \frac{m_{\widetilde{\mu}_{\mathrm{R}}}}{M_{1}}\right), \\
&\Delta a_{\mu}^{\mathrm{BHR}}=-\frac{g_{Y}^{2} m_{\mu}^{2}}{8 \pi^{2}} \frac{M_{1} \mu \tan \beta}{m_{\widetilde{\mu}_{\mathrm{R}}}^{4}} F_{b}\left(\frac{M_{1}}{m_{\widetilde{\mu}_{\mathrm{R}}}}, \frac{\mu}{m_{\widetilde{\mu}_{\mathrm{R}}}}\right), \\
&\Delta a_{\mu}^{\mathrm{BHL}}=\frac{g_{Y}^{2} m_{\mu}^{2}}{16 \pi^{2}} \frac{M_{1} \mu \tan \beta}{m_{\widetilde{\mu}_{\mathrm{L}}}^{4}}F_{b}\left(\frac{M_{1}}{m_{\widetilde{\mu}_{\mathrm{L}}}}, \frac{\mu}{m_{\tilde{\mu}_{\mathrm{L}}}}\right), 
\end{aligned}
\label{eqn:g2SUSY}
\end{equation}
where the loop {functions are} defined to be
\begin{equation}
\begin{aligned}
F_{a}(x, y)&=\frac{1}{2} \frac{C_{1}\left(x^{2}\right)-C_{1}\left(y^{2}\right)}{x^{2}-y^{2}},\quad F_{b}(x, y)=-\frac{1}{2} \frac{N_{2}\left(x^{2}\right)-N_{2}\left(y^{2}\right)}{x^{2}-y^{2}}, \\
C_1(x)&=\frac{3-4x+x^2+2\log x}{(1-x)^3},\quad N_2(x)=\frac{1-x^2+2x\log x}{(1-x)^3}. 
\end{aligned}
\end{equation}
From Eq.(\ref{eqn:g2SUSY}), light slepton mass is favored for explaining muon $(g-2)$ data. However light slepton is highly constrained by null result of SUSY search in LHC. We have to set $M_2$ to be $1\mathrm{TeV}$ and $m_{\tilde\mu_L}=m_{\tilde\mu_R}=m_{\tilde\nu}>700~\mathrm{GeV}$ to escape LHC exclusion limit \cite{Aad:2019vnb}.

	

Besides of the above contributions, there are also Barr-Zee two-loop contributions involving the sgoldstino in the diagrams. As a result, heavy slepton masses are still available for muon $(g-2)$ anomaly. Recall that, sgoldstino contains direct interaction with photon. The two-loop Barr-Zee diagram can be effectively regarded as one-loop \cite{Davoudiasl:2018fbb},
\begin{equation}
\Delta a^{\mathrm{Barr-Zee}}=\frac{m_{\mu}\lambda_{\gamma}\lambda_l}{4\pi^2}\mathcal{I}\left(\frac{m_{\mu}}{m_s}\right), 
\end{equation}
where $\lambda_{\gamma}=\sqrt{2}M_{\gamma}/F$. Combining both of contributions can yield suitable $\Delta a_{\mu}$.  Therefore, we can use $\Delta a_{\mu}$ to reduce the number of input parameters. For example $\sqrt{F}$ can be solved by imposing muon $(g-2)$ constraint. The SUSY and Barr-Zee contributions either can be competitive or be dominated by one of them in different parameter space. We select two benchmark points in Tab.~\ref{tab:benchmarks} which properly satisfy the $(g-2)_\mu$ anomaly (which also explain the KOTO anomaly as will be studied in the following). For the first one, SUSY and Barr-Zee diagrams are almost equally contributed to the discrepancy with ratio ${\Delta a_\mu^{\rm BZ}}/{\Delta a_\mu^{\rm SUSY}}=0.94$,    while  Barr-Zee diagram dominates the contribution in the second parameter set, as the ratio is 15.5.

\begin{table}[htbp]
\begin{centering}
\begin{tabular}{|c|c|c|c|c|c|c|} \hline  
Benchmarks     & 
$m_s$ (MeV)        &  
$\sqrt F$ (GeV)    & 
$\mu$ (TeV)        & 
$m_{\rm slepton}$ (GeV)  & 
${\Delta a_\mu^{\rm BZ}}/{\Delta a_\mu^{\rm SUSY}}$   \\ \hline   
BP1   & 
93    & 
3485  & 
15    & 
800   & 
0.94\\ \hline
BP2  & 
80    &
2880  & 
3     & 
1500  & 
15.5\\ \hline
\end{tabular}
\caption{Benchmarks for explaining both $(g-2)_\mu$ and KOTO anomalies. We set $A_0/\sqrt{F}=0.2, M_1=100 {\rm GeV}, M_2=1 {\rm TeV} $ for both cases, and $m_{\rm slepton}=m_{\tilde\mu_L}=m_{\tilde\mu_R}=m_{\tilde\nu}$. The benchmark with $m_s$ = 93 MeV is indicated in figures  by black point. The slepton masses chosen here are valid against the LHC constraints \cite{Aad:2019vnb}, which sets upper limits to be about 700 GeV with small nuetrilino mass. }
\label{tab:benchmarks} 
\end{centering}
\end{table}

Now we are in position to show how sgoldstino mimic the signal in the KOTO experiment.
{The effective branching ratio is }
\begin{equation}
\operatorname{BR}\left(K_{L} \rightarrow \pi^{0} s ; \mathrm{KOTO}\right)=\epsilon_{\mathrm{eff}} \mathrm{BR}\left(K_{L} \rightarrow \pi^{0} s\right) e^{-\frac{L}{p_s} \frac{m_s}{\tau_s}}. 
\label{eq:kotoBR}
\end{equation}
The efficiency factor $\epsilon_{\rm eff}$ can be read from \cite{Ahn:2018mvc}. $L$ is the detector size of KOTO experiment and chosen to be $3$ meters. The characteristic energy in KOTO experiment is $1.5~\mathrm{GeV}$. $\tau_{s}$ is the lifetime of sgoldstino.  In our sgoldstino model, we have $4$ input parameters in total: $A_0/\sqrt{F}, \sqrt{F}, m_s, M_1,M_2$. We should emphasize that $M_2$ is fixed to be $1~\mathrm{TeV}$ from LHC exclusion limit. The ratio $A_0/\sqrt{F}$ and $M_1$ is set to be $0.2$ and $100~\mathrm{GeV}$ respectively. As a result, we have only two parameters $m_s$ and $\sqrt{F}$ to testify the signal and constraints.

\begin{figure}[!htbp]
	\centering
	\includegraphics[width=0.65\textwidth]{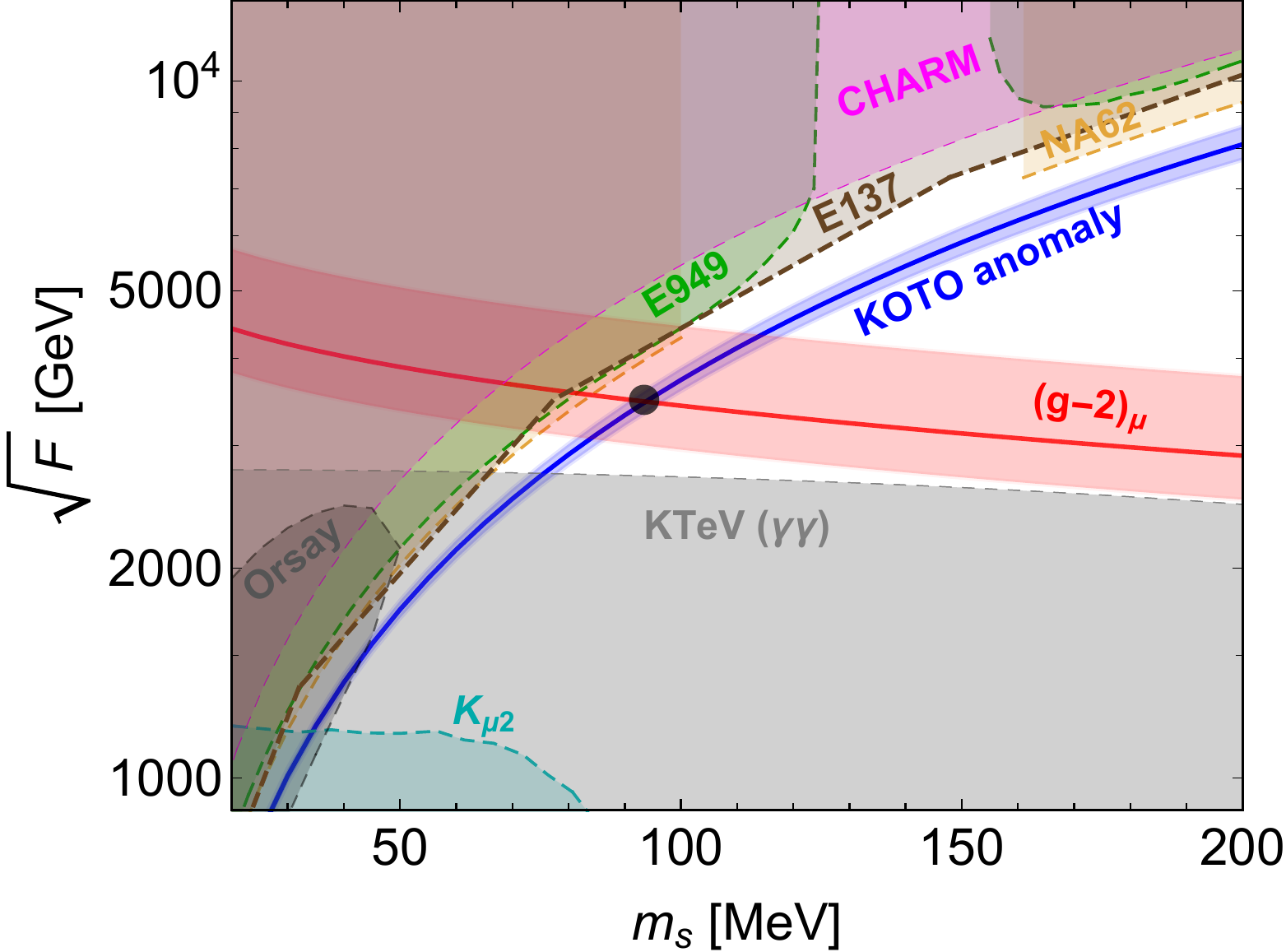}
	\caption{KOTO excess signals, $\mathrm{muon} (g-2)$ anomaly and beam-dump experimental constraints,
	plotted as functions of the scalar mass $m_s$ and SUSY breaking scale $\sqrt F$.
	Here we set $M_1=100$ GeV, $M_2=1$ TeV,
	$A_0/\sqrt F=0.2$, $m_{\tilde\mu_L}=m_{\tilde\mu_R}=m_{\tilde\nu}=800\mathrm{GeV}$, $\mu=15$ TeV.
	Blue: region in which can explain the KOTO anomaly.
	The blue line is the measured central value of the KOTO data.
	The blue band is the parameter space consistent with the KOTO
	anomalous events in 1 $\sigma$.
	Red: region of parameter space that can address the $(g-2)_\mu$ problem in $2\sigma$.
	The solid line corresponds to the central value,
	while the shaded regions {include the 2$\sigma$ compatible values}.
	Green: excluded regions  at $95\%$ C.L. by E949 experiment,
	which constraints on Br$(K^+\to\pi^+ X)$ with $X$ a long-lived particle.
	Light-Orange: limits from NA62 on Br$(K^+\to\pi^+\nu\bar\nu)$ at $95\%$ C.L..
	Magenta: limits on displaced decays of the scalar from the CHARM experiment,
	which measures displaced decay of neutral particles into $e^+e^-, \gamma\gamma, \mu^+\mu^-$.
	Black: the shaded region is excluded by the Orsay experiment
	which puts limits on the decaying of a scalar into electron pairs.
	Gray: limits from KTeV$(\gamma\gamma)$ on the process $K_L\to \pi^0\gamma\gamma$.
	Cyan: region excluded by $K_{\mu 2}$ from searching of light scalars in process $K^+\to\pi^+ s$.
	Brown: excluded regions from SLAC beam dump experiment E137 at $95\%$ C.L.
	The big black point in the parameters space corresponds to the benchmark
	that can explain both anomalies simultaneously.
	}
	\label{fig:final}
	\end{figure}

In figure \ref{fig:final},  we show our final results in space ($m_s - \sqrt{F}$) including both signals and constraints (here we use benchmark BP1 from Tab.\ref{tab:benchmarks} for illustration). In the blue band, we can obtain the KOTO signal events at the 95\% C.L., in which the solid blue curve stands for the central value of KOTO data. To explain the muon $g-2$ anomaly, we combine the contributions from {sgoldstino, neutralino and slepton one-loop diagrams, Barr-Zee diagrams}, then we find a red region in the parameter space where $\Delta a_\mu$ can be achieved {in $2 \sigma$ range}. We choose a benchmark point to satisfy both anomalies in figure shown by a black point, sgoldstino mass $m_s\sim90$ MeV and $\sqrt F\sim 3.5$ TeV.

Searching for rare decays of Kaons in a variety of beam dump experiments sets strong constraints on the light scalar. We demonstrate our analysis by showing the shaded excluded regions in the figure and discussing them in the following. 
\begin{itemize}
\item $\mathrm{E949}$ for green shaded region.\\
Measuring the decay $K^{+}\rightarrow \pi^{+}\nu\bar\nu$ by the $\mathrm{E949}$ collaboration has put constraints on the mimic  process $K^{+}\rightarrow \pi^{+} X$ with $X$ being long-lived particle. $\mathrm{E949}$ collaboration \cite{Artamonov:2009sz} explored the possibility of  such a process and provided the upper limits as function of scalar mass. We can thus use it to constrain our model easily with the effective  branching ratio in  Eq.\ref{eq:kotoBR}, except  the characteristic energy of $\mathrm{E949}$ is $0.71~\mathrm{GeV}$ and detector length is $4~\mathrm{m}$.

\item $\mathrm{NA62}$ for light-orange shaded region.\\
A similar constraint comes from the $\mathrm{NA62}$, which sets a 95\%C.L. bound on  the branching fraction of process  $K^{+}\rightarrow \pi^{+}\nu\bar\nu$,
 \begin{equation}
{\rm Br}(K^+\to\pi^+\nu\bar\nu)_{\rm NA62} < 2.44 \times10^{-10}.
\end{equation}
To apply the NA62 limits we also should use the effective branching ratio as used for KOTO in Eq.\ref{eq:kotoBR}. The NA62 detector size $L=150$ m; the scalar’s energy is taken to be approximately half of the charged kaon energy at this experiment, $E_s = 37$ GeV; for the NA62 effective branching fraction we set $\epsilon_{\rm eff} = 1$. NA62 did not constrain the $m_s$ in the mass range $[100~\mathrm{MeV}, \, 161~\mathrm{MeV}]$ because of large background in $K^{+}\rightarrow \pi^{+}\pi^0\rightarrow \pi^{+}\nu\bar\nu$. So we do not need to compute the bounds in this mass range. NA62 excludes the parameter space in the light-orange shaded regions, and a gap also shown as expected (see figure \ref{fig:final}).

 \item CHARM for magenta shaded region.\\
The CHARM experiment, which is a proton beam-dump experiment, measures the displaced decay of neutral particles into $\gamma\gamma$, $e^+e^-$ and $\mu^+\mu^-$ final states. Since our signal resulted from sgoldstino being produced from neutral and charged Kaon decay, then the sgoldstino decays into the $\gamma\gamma$, $e^+e^-$ final states, CHARM experiment is thus relevant for our model. The events number in the CHARM detector is \cite{Dolan:2014ska}
\begin{equation}
N_{\rm det}\simeq N_s (e^{-\frac{L_{\rm dump}}{c \tau_s} \frac{m_s}{p_s}} -e^{-\frac{L_{\rm dump}+L_{\rm fid}}{c \tau_s} \frac{m_s}{p_s}}),
\end{equation}
The exponential factors in determining the number of scalars that reach and decay within the detector volume. $L_{\rm dump} = 480$ m is the CHARM beam dump baseline, while $L_{\rm fid} = 35$ m is the detector fiducial length. The scalar momentum is obtained assuming an average scalar energy of $E_s = 12.5$ GeV \cite{Bergsma:1985qz}. $N_s=2.9\times 10^{17} \sigma_s/\sigma_{\pi^0}$ represents the number produced in the kaon decay, where $\sigma_s$ is the production cross section \cite{Bezrukov:2009yw},
\begin{equation}
\sigma_s \simeq \sigma_{pp} M_{pp} \chi_s (0.5  {\rm BR}(K^+\to\pi^+s)+0.25  {\rm BR}(K_L\to\pi^0s)),
\end{equation}
with $\sigma_{pp}$ the proton cross section, $M_{pp}$ is the total hadron multiplicity and $\chi_s = 1/7$ is the fraction of strange pair-production rate \cite{Andersson:1983ia}. For the neural pion yield  we have $\sigma_{\pi^0}\simeq\sigma_{pp} M_{pp}/3$. Due to the fact that CHARM experiment has observed zero event for such decays, we can set $90\%$ confidential level bound by requiring $N_{\mathrm{det}}<2.3$. The magenta shaded region in the parameter space has been excluded.

\item $\mathrm{K}_{\mu 2}$ for cyan shaded region.\\
The $K_{\mu 2}$ experiment \cite{Yamazaki:1984vg} searched for a neutral boson in a two-body decay of $K^+\to\pi^+ X$ with $X$ being the neutral scalar, and a momentum mono-chromatic $\pi^+$ was expected due to $K^+$ is stopped in the above 2-body decay. The null result of experiment thus set a constrain for our model. We translate the limits on our model parameters, as a result the cyan shaded region has been excluded as shown in figure \ref{fig:final}.

\item $\mathrm{KTeV}({\gamma\gamma}$) for gray shaded region.\\
The KTeV experiment is used to measure the process for neutral Kaon decay $K_L\rightarrow \pi^0\gamma\gamma$. The derived branching ratio for this process is \cite{Abouzaid:2008xm}
\begin{equation}
\mathrm{BR}(K_L\rightarrow \pi^0\gamma\gamma)=(1.29\pm 0.03\pm 0.05 )\times 10^{-6}.
\end{equation}
As a result, we can use this bound to constrain sgoldstino by conservatively setting the bound ${\rm BR}(K_L\rightarrow \pi^0 s){\rm BR} (s\to\gamma\gamma)<10^{-6}$\cite{Kitahara:2019lws,Liu:2020qgx}. Since the branching ratio of process $s\to\gamma\gamma$ closes to 1 for most ranges of $m_s$, the constraint is rather stringent, as shown by the gray region. Furhtermore, $\mathrm{KTeV}({e^+e^-})$ can also put a constraint to our model. But it highly depends on the branching ratio into electron pair for sgoldstino. From figure~\ref{fig:s-decay}, the photon final states dominates over electron. So we can safely ignore this constraint.
 
\item Orsay for black shaded region.\\
Orsay is an electron beam dump experiment which is sensitive to sgoldstino decaying into electron. It is similar with  $\mathrm{KTeV}({e^+e^-})$. We employ the method used in \cite{Davier:1989wz}, and place the limits on our parameter space at the 95\%C.L. As discussed about the $\mathrm{KTeV}({e^+e^-})$ constraints, the limits of Orsay is much less constraining, see the black shaded region in figure \ref{fig:final}.

 \item SLAC E137 for brown shaded region.\\
Another electron beam dump experiments, the E137 experiment \cite{Bjorken:1988as} at SLAC, has reported results from an analysis of axion coupling only to photons. The parameter space should be constrained by the lack of signal at E137 experiment, since scalar $s$ predominantly decays into photons in our case. Refs.\cite{Batell:2017kty, Dobrich:2015jyk, Dolan:2017osp} have studied the exclusion limits of a scalar coupled to photons through dimension-five operators. We translate these limits on the strength of the interaction $s F^{\mu\nu}F_{\mu\nu}$  into limits on $\sqrt{F}$ (see Eq.\ref{eqn:eff}), which is shown in brown color in figure \ref{fig:final}. The parameter space that addresses both KOTO anomaly and $(g-2)_\mu$ discrepancy is safe against these limits. 
\end{itemize}

\section{Conclusion}  \label{sec:conclusion}
 Supersymmetry is one of the most attractive new physics scenarios for solving hierarchy problem. Once the supersymmetry is broken, there exists a goldstino $\widetilde G$, and its superpartner is accordingly called sgoldstino $s$. Its mass can be light if the couplings induced Kahler potential are $\mathcal{O}(1)$.  In this work, we had explored the possibility that the sgoldstino can explain KOTO anomaly and muon $g-2$ simultaneously. The interactions between sgoldstino and quarks generate the flavor-changing neutral-current transition from strange quark to down quark via penguin diagrams. The resulting $K_L \to \pi^0 s$ transition followed by the decay of $s\to \gamma \gamma$ explains the KOTO signal. Although the coupling between lepton and sgoldstino is too small to contribute the desirable muon $g-2$, the contributions from neutralino and slepton at one-loop diagrams, Barr-Zee diagrams can explain this discrepancy.  We also studied all known experimental constraints such as from NA62, E949, E137, KOTO, Orsay, KTEV and CHARM experiments, and found that the mass of CP-even sgoldstino around pion mass can account for KOTO signal without violating the Grossman-Nir bound. In addition, we perform a comprehensive study with a benchmark point where sgoldstino mass is $m_s\sim 90$ MeV and SUSY breaking scale $\sqrt F\sim 3.5$ TeV. The parameter spaces can be further tested in future NA62, DUNE experiments, as well as experiments in the LHC for sleptons.

\begin{acknowledgments}
We especially thanks Xiaoping Wang for helpful discussions. This research was supported by the National Natural Science Foundation of China under the grants No. 11875062 and 11947302 (for TL),  11975195 (for YL), 11805161 (for BZ), 11947034 (for XL), by the Natural Science Foundation of Shandong Province under the grants No. ZR2019JQ004 (YL) and ZR2018QA007 (BZ), and by the Key Research Program of Frontier Science, CAS (TL). BZ is also supported by the Basic Science Research Program through the National Research Foundation of Korea (NRF) funded by the Ministry of Education, Science and Tech- nology (NRF-2019R1A2C2003738), and by the Korea Research Fellowship Program through the NRF funded by the Ministry of Science and ICT (2019H1D3A1A01070937). This work is  also supported by the Project of Shandong Province
Higher Educational Science and Technology Program under Grants No. 2019KJJ007.
\end{acknowledgments}

\bibliographystyle{apsrev4-1.bst}
\bibliography{lit}

\end{document}